\documentclass[nofootinbib,prd,aps,onecolumn,preprintnumbers,amsmath,amssymb,superscriptaddress,preprint]{revtex4}
\usepackage{graphicx}
\usepackage{latexsym}
\usepackage{threeparttable}
\usepackage{tabularx}
\usepackage{mathrsfs}
\usepackage[normalem]{ulem}        % Part of the standard distribution
\usepackage{multirow}
\usepackage{xcolor}
\usepackage{color}
\usepackage[colorlinks,linkcolor=magenta,anchorcolor=blue,citecolor=green]{hyperref}
\usepackage{pifont}

\allowdisplaybreaks[4]

\begin{document}

\tolerance=5000

\title{Nonpropagating ghost in covariant $f(Q)$ gravity}

\author{Kun~Hu}
\email[]{hukun@mails.ccnu.edu.cn}
\affiliation{Institute of Astrophysics, Central China Normal University, Wuhan 430079, China}

\author{Makishi~Yamakoshi}
\email[]{yamakoshi.makishi.b1@s.mail.nagoya-u.ac.jp}
\affiliation{
Department of Physics, Nagoya University, Nagoya 464-8602, Japan
}

\author{Taishi~Katsuragawa}
\email[(Corresponding author) ]{taishi@mail.ccnu.edu.cn}
\affiliation{Institute of Astrophysics, Central China Normal University, Wuhan 430079, China}

\author{Shin'ichi~Nojiri}
\email[]{nojiri@gravity.phys.nagoya-u.ac.jp}
\affiliation{
Department of Physics, Nagoya University, Nagoya 464-8602, Japan
}
\affiliation{
Kobayashi-Maskawa Institute for the Origin of Particles and the Universe, Nagoya University, Nagoya 464-8602, Japan
}

\author{Taotao~Qiu}
\email[]{qiutt@hust.edu.cn}
\affiliation{School of Physics, Huazhong University of Science and Technology, Wuhan, 430074, China}

\begin{abstract}
$f(Q)$ gravity is an extension of the symmetric teleparallel equivalent to general relativity (STEGR). 
This work shows that based on the scalar-nonmetricity formulation, a scalar mode in $f(Q)$ gravity has a negative kinetic energy.
This conclusion holds regardless of the coincident gauge frequently used in STEGR and $f(Q)$ gravity.
To study the scalar mode, 
we further consider the covariant $f(Q)$ gravity as a special class in higher-order scalar tensor (HOST) theory and rewrite the four scalar fields, which play a role of the St\"{u}eckelberg fields associated with the diffeomorphism, by vector fields.
Applying the standard Arnowitt-Deser-Misner (ADM) formulation to the new formulation of the $f(Q)$ gravity,
we demonstrate that the ghost scalar mode can be eliminated by the second-class constraints, thus ensuring that $f(Q)$ gravity is a healthy theory.
\end{abstract}

\maketitle

%%%%%%%%%%%%%%%%%%%%%%%%%
\section{Introduction}
%%%%%%%%%%%%%%%%%%%%%%%%%

In recent years, gravitational theories with nonmetricity have been actively discussed~\cite{Nester:1998mp,BeltranJimenez:2018vdo,BeltranJimenez:2019esp,Runkla:2018xrv,BeltranJimenez:2019tme}.
These theories are written by the nonmetricity scalar $Q$ as the fundamental geometrical quantity and in Palatini formalism, where the connection is an independent variable in addition to the metric.
Imposing that the Riemann tensor and torsion tensor vanish for the connection, one finds that the connection is written by the four scalar fields~\cite{Blixt:2023kyr,BeltranJimenez:2022azb,Adak:2018vzk,Tomonari:2023wcs}.
Moreover, one can choose specific scalar fields so that the connection vanishes, which is called the coincident gauge, and then the metric is only a dynamical variable in theory.
It is known that in the coincident gauge, the theory with action linear to $Q$ is equivalent to general relativity (GR), and we call it symmetric teleparallel equivalent to general relativity (STEGR)~\cite{BeltranJimenez:2017tkd,Quiros:2021eju}.

Teleparallelism is another well-known concept in the gravitational theory that does not rely on curvature, where the torsion scalar $T$ is the fundamental quantity~\cite{Maluf:2013gaa,Aldrovandi:2013wha,Cai:2015emx}.
By choosing the Weitzenb\"{o}ck connection, which corresponds to the coincident gauge in STEGR, the spin connection vanishes, and the tetrad is only a dynamical variable.
The theory with action linear to $T$ is equivalent to GR, and we call it teleparallel equivalent to general relativity (TEGR)~\cite{Maluf:2013gaa,Krssak:2015oua}.
Compared with teleparallelism, we utilize {\it symmetric} teleparallelism in nonmetricity gravity to reflect the torsionless condition, that is, symmetric affine connection.
Moreover, as in the way of extending GR to $f(R)$ gravity characterized by the function of the curvature scalar, TEGR, and STEGR have been extended to $f(T)$ and $f(Q)$ gravity whose actions include the arbitrary function of $T$ and $Q$ respectively~\cite{Bahamonde:2015zma,Cai:2015emx,Krssak:2015oua,Bahamonde:2021gfp,Heisenberg:2018vsk,BeltranJimenez:2019tme,Harko:2018gxr,Jarv:2018bgs,Runkla:2018xrv,Capozziello:2022zzh,Capozziello:2023vne}.
Those theories have been intensively examined as one of the modified gravity theories, 
and new degrees of freedom (DOF) introduced by the torsion or nonmetricity show various interesting phenomena;
for instance, cosmological models~\cite{Hu:2023ndc,Qiu:2018nle,Najera:2021afa,Li:2021mdp,Khyllep:2021pcu,Casalino:2020kdr,Sharma:2021fou,Sharma:2021ivo,Sharma:2021ayk,Mandal:2020buf,Mandal:2020lyq,Mandal:2021bpd,Arora:2022mlo,Gadbail:2022jco,Paliathanasis:2023kqs,Paliathanasis:2023nkb,Dimakis:2022rkd,Paliathanasis:2023gfq}, black hole solutions~\cite{Wang:2021zaz,Lin:2021uqa,DAmbrosio:2021zpm,Bahamonde:2022esv,Calza:2022mwt}, and gravitational waves~\cite{Hohmann:2018jso,Hohmann:2018wxu,Soudi:2018dhv,Abedi:2017jqx}.

Recently, the number of DOF in $f(Q)$ gravity has come into the spotlight~\cite{Hu:2022anq,DAmbrosio:2023asf,Tomonari:2023wcs,Heisenberg:2023lru,Paliathanasis:2023pqp,Dimakis:2021gby}.
In our previous work~\cite{Hu:2022anq}, we have shown there are eight DOFs in coincident $f(Q)$ gravity, 
and the scalar mode cannot propagate~\footnote{The conclusion in Ref.~\cite{Hu:2022anq} have been controversial and under active discussion recently. For example, see Refs.~\cite{DAmbrosio:2023asf,Tomonari:2023wcs} for details.}.
However, the existing works on the Hamiltonian analysis rely on the specific gauge, the so-called coincident gauge. 
It is significant to confirm that there are no ghost modes with negative kinetic energy in the physical DOF without gauge fixing.
This work mainly discusses the ghost scalar mode in the $f(Q)$ gravity.
Moreover, we investigate the ADM formulation of $f(Q)$ gravity in the arbitrary gauge to consider the origin of the negative kinetic energy.

This paper is organized as follows:
We introduce the underlying geometrical background of the $f(Q)$ gravity in Sec.~\ref{sec2}.
We apply the conformal rescaling to the action and discuss the ghost scalar mode in the $f(Q)$ gravity by the scalar-nonmetricity formulation in Sec.~\ref{sec3}. In Sec.~\ref{sec4}, we propose a new formulation of the $f(Q)$ gravity as the higher-order scalar-tensor theory. 
Finally, we conclude this paper and discuss the implications of ghost mode in Sec.~\ref{sec5}.
Some mathematical details of the paper are given in the Appendices.

Throughout this paper, we use the leading letters of the Latin alphabet $\left(a,b,c\right)$ running from $0$ to $3$ to label the tangent space-time coordinates. 
The Latin indices $\left(i,j,k\right)$ running from $1$ to $3$ represent the ADM spatial indices, and the Greek indices $\left( \alpha,\beta, \cdots \right)$ running from $0$ to $3$ does the space-time indices.
For clarity of notation, we define symbols as in Table~\ref{tab:notation}.
\begin{table}[htbp]\renewcommand\arraystretch{1.4}
    \caption{Conventions and notations}
    \label{tab:notation}
    \centering
    \begin{tabular}{|c|l|}
    \hline 
    $\left\{ {}^{\, \alpha}_{\mu \nu} \right\}$ & Levi-Civita connection \\
    $\varGamma ^{\alpha}_{\mu \nu}$& General affine Connection \\
    % $\left\{ {}^{\, i}_{\,j k} \right\}$ & Levi-Civita connection with respect to $h_{ij}$ \\
    $\nabla_{\mu}$ & Covariant derivative with respect to Levi-Civita connection \\
    ${\hat\nabla}_{\alpha}$ & Covariant derivative with respect to general affine connection. \\
     $\mathcal{D} _i$ & Three dimensional covariant derivative with respect to $h_{ij}$ \\
    % $\mathcal{L}_{\vec{N}}$ & Lie derivative with respect to the vector field \\
    $ R$ & Curvature scalar with respect to general affine connection \\
    $\mathcal{R}$ & Curvature scalar with respect to Levi-Civita connection \\
    $\mathring{Q}$ & Nonmetricity scalar in coincident gauge \\
    % ${}^3\mathcal{R}$ & Three dimensional curvature scalar with respect to Levi-Civita connection \\
    % ${}^3\mathcal{Q}$ & Three dimensional non-metricity scalar with respect to $h_{ij}$ \\
    \hline
    \end{tabular}
\end{table}

%%%%%%%%%%%%%%%%%%%%%%%%%
\section{Basics of $f(Q)$ Gravity and Coincident Gauge} \label{sec2}
%%%%%%%%%%%%%%%%%%%%%%%%%

We briefly review the basic structure of the $f(Q)$ gravity and symmetries. 
We discuss the four scalar fields in the affine connection and their role as the St\"{u}eckelberg fields.
%
%%%%%%%%%%%%%%%%%%%%%%%%%
\subsection{Geometrical foundations}
\label{sec2A}
%%%%%%%%%%%%%%%%%%%%%%%%%

A general affine connection $\varGamma^{\alpha}_{\,\mu \nu}$ can be decomposed into three parts:
\begin{align}
    \varGamma^{\alpha}_{\,\mu \nu}
    = \left\{ {}^{\, \alpha}_{\mu \nu} \right\} + K_{\,\mu \nu}^{\alpha} + L_{\,\mu \nu}^{\alpha}
    \, .
\end{align}
Here, $\left\{ {}^{\, \alpha}_{\mu \nu} \right\}$ is Levi-Civita connection
\begin{align}
    \left\{ {}^{\, \alpha}_{\mu \nu} \right\}
    = \frac{1}{2} g^{\alpha \lambda}\left(\partial_\mu g_{\lambda \nu} + \partial_{\nu} g_{\mu \lambda} - \partial_{\lambda}g_{\mu \nu} \right)
    \label{Levi-Civita connection}
    \, ,
\end{align}
$K_{\,\mu \nu}^{\alpha}$ is contortion
\begin{align}
    K^{\alpha}_{\,\mu \nu} 
    = \frac{1}{2} g^{\alpha \lambda} \left( T_{\mu\lambda \nu} + T_{ \nu\lambda\mu } + T_{\lambda\mu\nu } \right) 
    \, ,
\end{align}
and $L_{\,\mu \nu}^{\alpha}$ is disformation
\begin{align}
    L^{\alpha}_{\, \mu \nu} 
    = \frac{1}{2} g^{\alpha \lambda} \left(Q_{\lambda \mu \nu}- Q_{\mu \lambda \nu} - Q_{\nu \lambda \mu} \right)
    \, .
\end{align}
Torsion tensor $T^{\lambda}_{ \,\mu \nu}$ and nonmetricity tensor $Q_{\alpha \mu \nu}$ are defined as
\begin{align}
    T^{\alpha}_{\ \mu \nu} &\equiv  \varGamma^{\alpha}_{\ \mu\nu} - \varGamma^{\alpha}_{\ \nu \mu}
    \, , \\
    \begin{split}
    Q_{\alpha \mu \nu} 
    &\equiv \hat{\nabla}_{\alpha} g_{\mu \nu}
    \\
    &= \partial_{\alpha}g_{\mu\nu} - g_{\nu\sigma}\varGamma^{\sigma}_{\ \mu\alpha} - g_{\sigma\mu}\varGamma^{\sigma}_{\ \nu\alpha}
    \, .
    \end{split}
    \label{Eq: nonmetricity tensor}
\end{align}
Using two types of trace of nonmetricity tensor
\begin{align}
    Q_{\alpha} \equiv Q_{\alpha}{ }_{\mu}{ }^{\mu}
    \, , \quad 
    \tilde{Q}_{\alpha} \equiv Q^{\mu}_{\ \mu \alpha}
    \, ,
\end{align}
we define the nonmetricity conjugate 
\begin{align}
    P^{\alpha \mu \nu} =
    -\frac{1}{4} Q^{\alpha \mu \nu}+\frac{1}{2} Q^{(\mu \nu) \alpha}+\frac{1}{4}\left(Q^\alpha-\tilde{Q}^\alpha\right) g^{\mu \nu}-\frac{1}{4} g^{\alpha(\mu} Q^{\nu)}
    \, , 
\end{align}
and nonmetricity scalar 
\begin{align}
\begin{split}
    Q 
    &= Q_{\alpha \mu \nu} P^{\alpha \mu \nu}
    \\
    &= -\frac{1}{4}Q^{\alpha \mu \nu } Q_{\alpha \mu \nu } 
    + \frac{1}{2}  Q^{\nu \mu \alpha } Q_{\alpha \mu \nu }   
    + \frac{1}{4}  Q^{\alpha}Q_{\alpha}  
    - \frac{1}{2} \tilde{Q}^{\alpha }Q_{\alpha} 
    \label{Eq: nonmetricity scalar}
    \, . 
\end{split}
\end{align}

The action of STEGR is 
\begin{align}
    S = \int \mathrm{d}^{4} x\left(
    \sqrt{-g} Q_{\alpha \mu \nu} P^{\alpha \mu \nu} + \lambda_{\alpha}{ }^{\beta \mu \nu}R^{\alpha}{ }_{\beta \mu \nu}
    + \lambda_{a}{ }^{\mu \nu} T^{\alpha}_{\ \mu \nu} \right)
    \label{Eq: STEGR action}
    \, .
\end{align}
Variation of the action \eqref{Eq: STEGR action} with respect to Lagrange multipliers $\lambda_{\alpha}{ }^{\beta \mu \nu}$ and $\lambda_{a}{ }^{\mu \nu}$ generates the torsion-free condition
\begin{align}
    T^{\alpha}_{ \,\mu \nu} \stackrel{!}{=}0
    \label{Eq: torsion-free}
\end{align}
and vanishing curvature
\begin{align}
    {R}_{\ \beta \mu \nu}^{\alpha}(\varGamma)
    \equiv 
    \partial_{\mu} \varGamma^{\alpha}_{\ \nu \beta} - \partial_{\nu} \varGamma_{\ \mu \beta}^{\alpha}
    + \varGamma_{\ \mu \lambda}^{\alpha}\varGamma_{\ \nu \beta}^{\lambda}
    - \varGamma_{\ \nu \lambda}^{\alpha} \varGamma_{\ \mu \beta}^{\lambda}\stackrel{!}{=}0
    \label{Eq:inertial connection}
    \, .
\end{align}
The former demands the general affine connection to be symmetric to its lowered two indices, 
as the latter implies that the connection must have the form $\varGamma^{\lambda}_{\,\mu \nu} = (A^{-1})^{\lambda}_{\ \alpha} \partial_{\mu}A^{\alpha}_{\ \nu}$~\cite{Zhao:2021zab}. 

%%%%%%%%%%%%%%%%%%%%%%%%%
\subsection{Coincident gauge} 
\label{subsecB}
%%%%%%%%%%%%%%%%%%%%%%%%%

Combining two conditions on the affine connection as in Eqs.~\eqref{Eq: torsion-free} and \eqref{Eq:inertial connection}, 
we find that the matrix $A^\alpha{ }_\beta=\partial_\beta \xi^\alpha$, and the affine connection consequently takes the following form:
\begin{align}
    \varGamma^{\lambda}_{\,\mu \nu}
    = \frac{\partial x^{\lambda}}{\partial \xi^{\alpha}} \frac{\partial^2 \xi^{\alpha}}{\partial x^\mu \partial x^{\nu}} 
    \label{Eq: connection1}
    \, .
\end{align}
Here, four scalar fields $\xi^{\alpha}(x)$ are the arbitrary functions of coordinates.
Note that the connection is symmetric with respect to two lower indices because two partial derivatives of $\xi^{\alpha}$ commute each other.
Moreover, one finds $A^{\alpha}_{\mu}$ corresponds to the tetrad $e^{a}_{\mu}$ for Weitzenb\"{o}ck connection in the teleparallel gravity.

A special choice $\xi^{\alpha}(x) = x^{\alpha}$ makes the connection vanish $\varGamma^{\lambda}_{\ \mu\nu}=0$.
This choice is called the coincident gauge. 
Under the coincident gauge, the covariant derivative is replaced by the partial derivative $\nabla _{\alpha} \mapsto \partial _{\alpha}$, 
and the nonmetricity tensor $Q_{\alpha \mu \nu} $, deformation tensor $L^{\alpha}_{\, \mu \nu}$, nonmetricity scalar $Q$, 
and the traces $Q_{\alpha}$, $\tilde{Q}_{\alpha}$ reduce to
\begin{align}
    Q_{\alpha \mu \nu} 
    \mapsto
    \mathring{Q}_{\alpha \mu \nu} 
    &= 
    \partial_{\alpha} g_{\mu \nu}
    \, , \\
    L^{\alpha}_{\, \mu \nu}
    \mapsto
    \mathring{L}_{\,\mu \nu}^{\alpha} 
    &= 
    -\left\{ {}^{\, \alpha}_{\mu \nu} \right\}
    \, , \\
    \begin{split}
    Q
    \mapsto
    \mathring{Q} 
    &= 
    g^{\mu \nu}\left(
    \mathring{L}_{\ \sigma \mu}^{\alpha} \mathring{L}_{\ \nu \alpha}^{\sigma}
    - \mathring{L}_{\ \sigma \alpha}^{\alpha} \mathring{L}_{\ \mu \nu}^{\sigma}
    \right)
    \\
    &= 
    g^{\mu \nu}\left(
    \left\{ {}^{\, \alpha}_{\sigma \mu} \right\} 
    \left\{ {}^{\, \sigma}_{\nu \alpha}  \right\} 
    -  \left\{ {}^{\, \alpha}_{\sigma \alpha} \right\} 
    \left\{ {}^{\, \sigma}_{\mu \nu} \right\} 
    \right) \, ,
    \end{split}
    \\
    \begin{split}
    Q^{\alpha} - \tilde{Q}^{\alpha} 
    \mapsto
    \mathring{Q}^{\alpha}-\mathring{\tilde{Q}}^{\alpha}
    &= 
    g^{\mu \nu} \mathring{L}^{\,\alpha}_{\,\mu\nu} 
    - g^{\mu \alpha} \mathring{L}^{\,\nu}_{\,\mu\nu} 
    \\
    &= 
    - g^{\mu \nu}\left\{ {}^{\,\alpha}_{\mu\nu} \right\}
    + g^{\mu \alpha}\left\{ {}^{\,\nu}_{\mu\nu} \right\}
    \, .
    \end{split}
    \label{Eq: coincident relation}
\end{align}
Writing the Einstein-Hilbert action as follows
\begin{align}
\begin{split}
    \mathcal{S}_{{\mathrm{EH}}}
    & = 
    \int d^{4}x \,\sqrt{-g} \, g^{\mu \nu}\mathcal{R}_{\mu\nu} \left(\left\{ {}^{\, \alpha}_{\mu \nu} \right\} \right)
    \\
    &= 
    \int d^{4}x \,\sqrt{-g} g^{\mu \nu}\left(
    \left\{ {}^{\,\alpha}_{\sigma \mu} \right\} \left\{ {}^{\,\sigma}_{\nu \alpha} \right\}
    -\left\{ {}^{\,\alpha}_{\sigma \alpha} \right\} \left\{ {}^{\,\sigma}_{\mu \nu} \right\}
    \right)
    \\
    & \quad
    + \int d^{4}x \,\partial _{\alpha}\left[ 
    \sqrt{-g} \,(g^{\mu \nu}\left\{ {}^{\,\alpha}_{\mu\nu} \right\}-g^{\mu \alpha}\left\{ {}^{\,\nu}_{\mu\nu} \right\}) \right] 
    \label{Eq: Hilbert-Einstein action2}
    \,
\end{split}
\end{align}
we find that the Ricci scalar can be expressed by the nonmetricity scalar and divergence of two independent traces of the nonmetricity tensor, where
the divergence term is reduced to the boundary term after the integration.
This result indicates the equivalence between GR and STEGR up to the boundary term.
Moreover, for general connection, we find the relation between $\mathcal{R}$:
\begin{align}
    \mathcal{R}
    = 
    Q - {\nabla}_{\alpha}\left(Q^{\alpha}-\tilde{Q}^{\alpha}\right)
    \label{relation.Q.R}
\end{align}

%%%%%%%%%%%%%%%%%%%%%%%%%
\subsection{Symmetry and St\"{u}eckelberg field} 
%%%%%%%%%%%%%%%%%%%%%%%%%

Under the general coordinate transformation $\{X^{\mu} \} \rightarrow \{x^{\mu} \}$, the connection is transformed as
\begin{align}
    \varGamma^{\lambda}_{\ \mu \nu} (x)
    =
    \frac{\partial x^{\lambda}}{\partial X^{\alpha}} \frac{\partial X^{\rho}}{\partial x^{\mu}} \frac{\partial X^{\sigma}}{\partial x^{\nu}} 
    \varGamma^{\alpha}_{\ \rho \sigma} (X)
    + \frac{\partial x^{\lambda}}{\partial X^{\alpha}} \frac{\partial^{2} X^{\alpha}}{\partial x^{\mu} \partial x^{\nu}}~.
    \label{Eq: connection trans}
\end{align}
Provided that the connection vanishes in the coordinate system $\{X^{\mu} \}$, 
the first term vanishes in Eq.~\eqref{Eq: connection trans}, and the second term reproduces Eq.~\eqref{Eq: connection1}.
Thus, we can identify a set of scalar fields $\{\xi^{\mu} \}$ as such a special coordinate system $\{X^{\mu}\}$.
Note that the above argument partially includes the definition of the local Lorentz frame, although the space-time is not necessarily flat in the current setup.

Equation~\eqref{Eq: connection1} also suggests that the index of the scalar fields is contracted within the connection, and it does not show up as a free index.
Therefore, one can expect that the index $\alpha$ in $\xi^{\alpha}$ is related to the inertial symmetry rather than the space-time index.
To distinguish it from the space-time index, hereafter, we denote the scalar field as $\xi^{a}$:
\begin{align}
    \varGamma^{\lambda}_{\,\mu \nu} = \frac{\partial x^{\lambda}}{\partial \xi^a} \frac{\partial^2 \xi^a}{\partial x^\mu \partial x^\nu}
    \label{Eq: connection2}
    \, . 
\end{align}
We can find that Eq.~\eqref{Eq: connection2} is invariant under the following transformation for $\xi^{a}$:
\begin{align}
    \xi^a (x) \rightarrow \bar{\xi}^a (x) = \mathcal{M}^{a}_{\ b} \xi^b(x) + \zeta^a
    \label{Eq: poincare trans}
    \, .
\end{align}	
Here, $\mathcal{M}^{a}_{\ b}$ is a $4 \times 4$ nondegenerate constant matrix, and $\zeta^a$ is a constant vector.
In the coincident gauge $\xi^{a} = x^{a}$, the above transformation leads to the linear transformation between two coordinate systems,
\begin{align}
    \bar{x}^a = \mathcal{M}^{a}_{\ b} x^b + \zeta^a
\label{Eq: poincare trans2}
\, .
\end{align}
Equation~\eqref{Eq: poincare trans2} represents the affine transformation, and the connection \eqref{Eq: connection2} is purely inertial, which always can transform to zero under Eq.~\eqref{Eq: poincare trans2}.
By imposing on the flat space-time, the Minkowski metric is invariant under the transformation, 
we find that matrix $\mathcal{M}^{a}_{\ b}$ corresponds to the Lorentz transformation,
and Eq.~\eqref{Eq: poincare trans2} is reduced to the Poincar\'{e} transformation in the coincident gauge.

Thus, any two different frames under the coincident gauge are related by transformation \eqref{Eq: poincare trans2}, naturally to introduce the metric $f_{ab}(x)$ in the coincident gauge. 
Under the coordinate transformation $\{\xi^{a} \} \rightarrow \{x^{\mu} \}$, the space-time metric is transformed as $f_{ab}(x) \rightarrow g_{\mu\nu}(x)$
written in terms of the Jacobi matrix: 
\begin{align}
    g_{\mu\nu} = \frac{\partial\xi^{a}}{\partial x^{\mu}}\frac{\partial\xi^{b}}{\partial x^{\nu}}f_{ab} 
    \label{Eq: general covariance}
    \, .
\end{align}
We present an explicit example in Appendix~\ref{appendix A} to better demonstrate how $f_{ab}$ transforms under Eq.~\eqref{Eq: general covariance}. 
It is apparent that $\xi^{a}$ plays a role in restoring the general covariance while fixing $\xi^a$ breaks the diffeomorphism explicitly.
A similar structure can be found in the relation between the tetrad and metric if we read $\partial_{\mu} \xi^{a} = e^{a}_{\mu}$ and $f_{ab} = \eta_{ab}$.
Moreover, Eq.~\eqref{Eq: general covariance} shows up in the mass term of the de Rham-Gabadadze-Tolley (dRGT) massive gravity~\cite{deRham:2010kj,deRham:2011rn,Hassan:2011vm,Hinterbichler:2011tt,Molaee:2017enn}
, and $\xi^a = x^{a}$ is called the unitary gauge therein.
The above arguments suggest that we can treat four scalar fields $\xi^{a}$ as the St\"{u}eckelberg fields associated with the diffeomorphism as discussed in the dRGT massive gravity.
Hereafter we will handle the generic $\xi^{a}$ to investigate the $f(Q)$ gravity in a covariant way.

%%%%%%%%%%%%%%%%%%%%%%%%%
\section{Ghost scalar field in $f(Q)$ gravity} \label{sec3}
%%%%%%%%%%%%%%%%%%%%%%%%%
In this section, we recall the scalar-nonmetricity formulation of $f(Q)$ gravity and apply the conformal re-scaling transformation to the action. 
We suggest that the scalar field is a ghost mode in the scalar-nonmetricity formulation.

%%%%%%%%%%%%%%%%%%%%%%%%%
\subsection{$f(Q)$ gravity and scalar-nonmetricity formulation}
%%%%%%%%%%%%%%%%%%%%%%%%%

Let us consider the $f(Q)$ gravity by replacing $Q$ with a function of $Q$:
\begin{align}
    S_{f(Q)}=\int d^{4} x \sqrt{-g} f(Q)
    \label{Eq: fQ}
    \, .
\end{align}
By introducing an auxiliary scalar field $\phi$, one can consider the following action
\begin{align}
    S_{f(Q)}= \int d^{4} x \sqrt{-g}\left\{ f^{\prime}(\phi) Q- \left[\phi f^{\prime}(\phi)-f(\phi) \right] \right \}\; .
\end{align}
By varying the above action with respect to $\phi$, one gets $f''(\phi)(Q - \phi) = 0$. 
Provided that $f''(\phi) \neq 0$, this equation implies $\phi= Q$ and restores the original action.
Moreover, we can redefine the scalar field $\varphi \equiv f'(\phi)$ and $V(\varphi)=\phi f^{\prime}(\phi)-f(\phi)$. 
The Lagrangian density of our interest then takes the following form
\begin{align}
    S_{f(Q)}= \int d^{4} x \sqrt{-g} \left[\varphi Q - V(\varphi) \right]~.
    \label{Eq: scalarnonmetricity}
\end{align}

Using Eq.~\eqref{relation.Q.R}, we obtain
\begin{align}
    S_{f(Q)} 
    = \int d^{4} x \sqrt{-g} \left[\varphi \mathcal{R} - V(\varphi) + \varphi \nabla_{\mu}\left(Q^{\mu}-\tilde{Q}^{\mu} \right) \right]~.
    \label{Eq: scalarnonmetricity2}
\end{align}
The third term includes the covariant derivative with respect to the Levi-Civita connection $\nabla_{\mu}$,
and we can utilize the ordinary formula for the divergence,
\begin{align}
    \int d^{4} x \sqrt{-g}~\nabla_{\mu} A^{\mu} = \int d^{4} x~\partial_{\mu} \left( \sqrt{-g} A^{\mu} \right) 
    \, ,
\end{align}
and ignore the surface integration.
The integration by parts reduces the original action \eqref{Eq: fQ} to the following form:
\begin{align}
    S_{f(Q)} 
    = \int d^{4} x \sqrt{-g} \left[\varphi \mathcal{R} - V(\varphi) - \partial_{\mu} \varphi \cdot \left(Q^{\mu}-\tilde{Q}^{\mu}\right) \right]~.
    \label{Eq: scalarnonmetricity3}
\end{align}

%%%%%%%%%%%%%%%%%%%%%%%%%
\subsection{Conformal rescaling of $f(Q)$ gravity} 
%%%%%%%%%%%%%%%%%%%%%%%%%

We further deform Eq.~\eqref{Eq: scalarnonmetricity3}, following the frame transformation established in $f(R)$ gravity.
We consider the following conformal transformation of the metric:
\begin{align}
    g_{\mu \nu} \rightarrow g^\prime_{\mu \nu} = e^{- \Phi} g_{\mu \nu}
    \, , \quad
    \Phi = - \ln \varphi~.~\label{Eq.conformal transformation}
\end{align}
\noindent
It is worth mentioning that the affine connection is invariant under the transformation~\eqref{Eq.conformal transformation} 
since the connection is written in Eq.~\eqref{Eq: connection2} by the St\"{u}eckelberg field, which is independent of the metric.
By the conformal transformation of metric, the Ricci scalar $\mathcal{R}$ with respect to the Levi-Civita connection is transformed as
\begin{align}
\mathcal{R} 
=&
\mathrm{e}^{-\Phi}
\left[
\mathcal{R}^{\prime}
- 3 g^{\prime \mu \nu} \nabla^{\prime}_{\mu} \partial_{\nu} \Phi 
- \frac{3}{2} g^{\prime \mu \nu} \left( \partial_{\mu} \Phi \right) \left( \partial_{\nu} \Phi \right) 
\right]
\label{Eq: ricci scalar trans}
\, .
\end{align}
Here, we denote by $\nabla^{\prime}$ the covariant derivative with respect to the Levi-Civita connection after the transformation.

Moreover, by the transformation, the nonmetricity tensor $Q_{\alpha \beta \gamma}$ and its traces $Q^{\mu}$ and $\tilde{Q}^{\mu}$ are transformed as below:
\begin{align}
    \begin{split}
        Q_{\alpha \beta \gamma} \rightarrow Q^\prime_{\alpha \beta \gamma} 
        & = \hat{\nabla}_\alpha g^\prime_{\beta \gamma} 
        \\
        & = \hat{\nabla}_\alpha \left(e^{- \Phi} g_{\beta \gamma} \right) 
        \\
        & = e^{- \Phi} Q_{\alpha \beta \gamma} - g_{\beta \gamma} e^{- \Phi} \partial_\alpha \Phi \; ,
    \end{split}
    \label{conformal trans. of the nonmetricity tensor1}
\end{align}
\begin{align}
    \begin{split}
        Q^{\mu} \rightarrow Q^{\prime \mu} 
        & = g^{\prime \mu \alpha} g^{\prime \beta \gamma} Q^\prime_{\alpha \beta \gamma}
        \\
        & = e^{2 \Phi} g^{\mu \alpha} g^{\beta \gamma} \left(e^{- \Phi} Q_{\alpha \beta \gamma} - g_{\beta \gamma} e^{- \Phi} \partial_\alpha \Phi \right) 
        \\
        & = e^\Phi Q^\mu - 4 e^\Phi g^{\mu \alpha} \partial_\alpha \Phi \; ,
    \end{split}
    \label{conformal trans. of the nonmetricity tensor2}
\end{align}
\begin{align}
    \begin{split}
        \tilde{Q}^{\mu} \rightarrow \tilde{Q}^{\prime \mu} 
        & = g^{\prime \mu \alpha} g^{\prime \beta \gamma} Q^\prime_{\beta \alpha \gamma} 
        \\
        & = e^{2 \Phi} g^{\mu \alpha} g^{\beta \gamma} \left(e^{- \Phi} Q_{\beta \alpha \gamma} - g_{\alpha \gamma} e^{- \Phi} \partial_\beta \Phi \right) 
        \\
        &= e^\Phi \tilde{Q}^\mu - e^\Phi g^{\mu \alpha} \partial_\alpha \Phi\; .
    \end{split}
\label{conformal trans. of the nonmetricity tensor3}
\end{align}
In Eq.~\eqref{conformal trans. of the nonmetricity tensor1}, the general affine connection in the covariant derivative $\hat{\nabla}$ are not transformed.
From the above, we can rewrite the original nonmetricity tensor $Q$ by new ones $Q^{\prime}$ and scalar field $\Phi$ after the conformal transformation,
\begin{align}
    \begin{split}
        Q^{\mu} 
        &= e^{- \Phi} Q^{\prime \mu} + 4 g^{\mu \alpha} \partial_\alpha \Phi 
        \\
        &= e^{- \Phi} Q^{\prime \mu} + 4 e^{-\Phi} g^{\prime \mu \alpha} \partial_\alpha \Phi \; ,
    \end{split}
\end{align}
\begin{align}
    \begin{split}
        \tilde{Q}^{\mu} 
        &= e^{- \Phi} \tilde{Q}^{\prime \mu} + g^{\mu \alpha} \partial_\alpha \Phi 
        \\
        &= e^{- \Phi} \tilde{Q}^{\prime \mu} + e^{-\Phi} g^{\prime \mu \alpha} \partial_\alpha \Phi\; .
    \end{split}
\end{align}
Thus, the third term in Eq.~\eqref{Eq: scalarnonmetricity3} is given as
\begin{align}
    \begin{split}
        \partial_\mu \varphi \cdot \left( Q^\mu -\tilde{Q}^\mu \right) 
        &= - e^{- \Phi} \partial_\mu \Phi \cdot e^{- \Phi} 
        \left(
        Q^{\prime \mu} + 4 g^{\prime \mu \alpha} \partial_\alpha \Phi - \tilde{Q}^{\prime \mu} -  g^{\prime \mu \alpha} \partial_\alpha \Phi 
        \right) \\
        &= - e^{-2 \Phi} \left[
        \partial_\mu \Phi \cdot \left(Q^{\prime \mu} - \tilde{Q}^{\prime \mu} \right) 
        + 3 \partial^\alpha \Phi \partial_\alpha \Phi \right]\; .
    \end{split}
    \label{Eq: boundary trans}
\end{align}
By substituting Eqs.~\eqref{Eq: ricci scalar trans} and \eqref{Eq: boundary trans} into Eq.~\eqref{Eq: scalarnonmetricity3}, the action of $f(Q)$ gravity is written by the Einstein gravity with the minimally coupled scalar field:
\begin{align}
    \begin{split}
        S 
        &= \int d^{4} x \sqrt{- g^\prime} e^{2 \Phi} 
        \left \{ \mathrm{e}^{-2\Phi} \left[
        \mathcal{R}^{\prime}
        - 3 g^{\prime \mu \nu} \nabla^{\prime}_{\mu} \partial_{\nu} \Phi 
        - \frac{3}{2} g^{\prime \mu \nu} \left( \partial_{\mu} \Phi \right) \left( \partial_{\nu} \Phi \right) 
        \right] 
        \right.
        \\
        & \qquad \qquad \qquad \qquad \qquad \left. 
        - V(\Phi) 
        + e^{-2 \Phi} \left[
        \partial_\mu \Phi \cdot \left(Q^{\prime \mu} - \tilde{Q}^{\prime \mu} \right) 
        + 3 \partial^\alpha \Phi \partial_\alpha \Phi \right] 
        \right \}
        \\
        &= \int d^4 x \sqrt{- g^\prime} \, \left[\mathcal{R}^\prime + \frac{3}{2} \partial^\alpha \Phi \partial_\alpha \Phi - U(\Phi) 
        + \partial_\mu \Phi \cdot \left(Q^{\prime \mu} - \tilde{Q}^{\prime \mu} \right) \right]\; .
    \end{split}
\label{action through conformal trans.}
\end{align}
Here, we defined $U(\Phi) \equiv e^{2 \Phi} V(\Phi)$ and ignored the divergence term.

In Eq.~\eqref{action through conformal trans.}, the sign of the kinetic term of scalar field $\Phi$ is positive, indicating that it is ghost mode.
The conformal transformation of the first and second terms in Eq.~\eqref{Eq: scalarnonmetricity3} reproduces the minimally coupled scalar field $\Phi$, 
which is the well-known result in $f(R)$ gravity.
However, since the third term, the nonmetricity tensor, in Eq.~\eqref{Eq: scalarnonmetricity3} gives rise to a new kinetic term with the opposite sign, the sign of the kinetic term of the scalar field is finally reversed.

As a classical theory, the ghost fields have negative kinetic energy and generate the instability of the system. 
In a quantum context, the energy is bounded below by the vacuum, even if there is a ghost. 
The ghosts, however, generate negative norm states, which give the negative probability; therefore, the existence of the ghosts conflicts with the Copenhagen interpretation of the quantum theory, and the theory is not physically acceptable.
In the case of the gauge theory, if we quantize the system by using the BRS symmetry \cite{Becchi:1975nq}, the ghosts appear in the combinations of zero norm states in the physical states, which satisfy the constraints coming from the BRS symmetry or gauge symmetry, and therefore the negative norm states are eliminated \cite{Kugo:1977zq, Kugo:1979gm}. 
The situation in $f(Q)$ gravity could be similar. 
Although there appear to be ghosts even in $f(Q)$ gravity, the propagation of the ghost fields is excluded as the physical degrees of freedom by the constraints, as we will see in the next section.

%%%%%%%%%%%%%%%%%%%%%%%%%
\section{Reformulation of $f(Q)$ gravity} \label{sec4}
%%%%%%%%%%%%%%%%%%%%%%%%%
In this section, we reformulate the action of $f(Q)$ gravity.
Regarding the $f(Q)$ gravity as a higher-derivative scalar-tensor theories~\cite{Belenchia:2016bvb,Langlois:2015skt,Gao:2018izs,Kimura:2016rzw,Crisostomi:2016czh,BenAchour:2016cay}, we develop ADM formulation of the covariant action and discuss the origin of the ghost scalar mode.

\subsection{Scalar-vector-tensor formulation}

The nonmetricity scalar $Q$ includes the metric and its first derivative as in Eqs.~\eqref{Eq: nonmetricity tensor} and \eqref{Eq: nonmetricity scalar}, 
while the connection includes the first and second derivatives of St\"{u}eckelberg fields as in Eq.~\eqref{Eq: connection2}. 
Thus, the Lagrangian density of $f(Q)$ gravity~\eqref{Eq: scalarnonmetricity} is symbolically given as
\begin{align}
    \sqrt{-g} \mathcal{L} \, \left(\partial_{\nu}\xi^{a},\partial_{\mu}\partial_{\nu}\xi^{a},g_{\mu\nu},\partial_{\beta}g_{\mu\nu}\right)
    \label{second derivative lagrange}
    \, .
\end{align}
To handle the higher-order derivatives in the Lagrangian density, we can introduce new variables into the $f(Q)$ theory in a way that does not alter the theory~\cite{Motohashi:2016ftl}. 
By introducing four space-time vectors $A^a_{\;\mu}$ as new variables, we can rewrite the Lagrangian density~\eqref{second derivative lagrange} as~\footnote{
Taking the variation of~\eqref{eq.action} with respect to $\omega_{\;a}^{\nu}$, we can get $\partial_{\nu}\xi^{a}=A_{\;\nu}^{a}$.
By inserting this relation back into the equation of motion, it is easy to check that the Lagrangian~\eqref{eq.action} is equivalent to~\eqref{second derivative lagrange}.
}
\begin{align}
    \sqrt{-g} \mathcal{L}^{eq}
    =\sqrt{-g} \mathcal{L}\, (A^{a}_{\; \nu}, \partial_{\mu}A^{a}_{\; \nu}, g_{\mu\nu}, \partial_{\beta}g_{\mu\nu})
    + \omega_{\;a}^{\nu}(\partial_{\nu}\xi^{a}-A_{\;\nu}^{a})
    \label{eq.action}
    \, .
\end{align}
We note that by rewriting $\partial_{\nu}\xi^{a}=A_{\;\nu}^{a}$, the affine connection Eq.~\eqref{Eq: connection2} goes back the original form suggested by Eq.~\eqref{Eq:inertial connection}.

Using Eq.~\eqref{eq.action}, we can reformulation the action Eq.~\eqref{Eq: scalarnonmetricity3} into following form~\footnote{We present the detail calculation in Appendix~\ref{appendix B}},
\begin{align}
    S_{f(Q)}^{eq} 
    % & =\int d^{4}x\,  \left[ \sqrt{-g}\left(\varphi Q - V \right)
    % + \omega_{\;a}^{\nu}(\partial_{\nu}\xi^{a}-A_{\;\nu}^{a}) \right]
    % \nonumber \\
    & =
    \int d^{4}x\, \left[ \sqrt{-g} \,
    \left(\varphi \mathcal{R} - V \right) - \sqrt{-g} \partial_{\alpha} \varphi \cdot \left(Q^{\alpha} - \tilde{Q}^{\alpha} \right) 
    + \omega_{\;a}^{\nu}(\partial_{\nu}\xi^{a}-A_{\;\nu}^{a})\right]
    \nonumber \\
    & =
    S_{f(R)} - \int d^{4}x\, \left \{ 
    \sqrt{-g} \, \partial_{\alpha}\varphi \left[
    g^{\mu\nu}(A^{-1})^{\alpha}_{\ a}
    - g^{\mu\alpha}(A^{-1})^{\nu}_{\ a}\right]\cdot\nabla_{\mu}A^{a}_{\ \nu}
    - \omega^{\nu}_{\ a} \left(\partial_{\nu}\xi^{a} - A^{a}_{\ \nu} \right) 
    \right \}
    \nonumber \\
    & =
    S_{f(R)}
    - \int d^{4}x\, \left[
    \sqrt{-g} \, C_{a}^{\alpha\mu\nu} \partial_{\alpha}\varphi \nabla_{\mu}A^{a}_{\ \nu}
    - \omega^{\nu}_{\ a} \left(\partial_{\nu}\xi^{a} - A^{a}_{\ \nu} \right) 
    \right]
    \label{Eq.modified action}
    \, .
\end{align}
To simplify the expression in Eq.~\eqref{Eq.modified action}, we defined a tensor $C_{a}^{\alpha\mu\nu}$ as
\begin{align}
    C_{a}^{\alpha\mu\nu} \equiv g^{\mu\nu}B^{\alpha}_{\ a} - g^{\mu\alpha}B^{\nu}_{\ a}
    \quad \text{where} \quad 
    B^{\alpha}_{\ a}=(A^{-1})^{\alpha}_{\ a} 
    \, .   \label{Eq.tensor C}
\end{align}
Where the term $S_{f(R)}$ in action~\eqref{Eq.modified action} is the scalar-tensor description of $f(R)$ gravity,
\begin{align}
    S_{f(R)} = \int d^{4}x\, \sqrt{-g} \left[\varphi \mathcal{R} - V(\varphi) \right]
    \, .
\end{align}

We note that the nonmetricity tensor in Eq.~\eqref{Eq: scalarnonmetricity3} induces the coupling between the scalar field $\varphi$ and vector field $A^{a}_{\ \mu}$ in Eq.~\eqref{Eq.modified action}.
This term corresponds to the origin of the ghost scalar mode after the conformal transformation of metric, as we observed in the previous section.

%%%%%%%%%%%%%%%%%%%%%%%%%
\subsection{$3+1$ decomposition of the action } 
\label{subsec}
%%%%%%%%%%%%%%%%%%%%%%%%%

In this subsection, we investigate the origin of the ghost mode under the Lagrangian formulation by performing the $3+1$ decomposition of the action~\eqref{Eq.modified action} in terms of ADM variables.
We decompose space-time into 3-dimensional spacelike hypersurfaces $\Sigma_{t}$ and their normal vector $n^{\mu}$, which satisfies the normalization condition $n^{\mu}n_{\mu}=-1$
\begin{align}
    n_{\mu} 
    &=
    \left(-N, 0, 0, 0 \right) 
    \, , \ 
    n^{\mu} 
    =
    \left( 1/N, -N^{i}/N \right)
    \, .
\end{align}
$ N$ is the lapse function and $N^i$ is the shift vector lying on the $\Sigma_{t}$. We also introduce the three-dimensional induced metric $h_{\mu\nu}$ defined by
\begin{align}
\begin{split}
    h_{\mu \nu} 
    &=
    g_{\mu \nu}+n_{\mu} n_{\nu} 
    = \left(
        \begin{array}{cc}
        N_{i} N^{i} & N_{i} \\
        N_{i} & h_{i j}
        \end{array}
        \right)
    \, , \\
    h^{\mu \nu} 
    & = 
    g^{\mu \nu}+n^{\mu} n^{\nu} 
    = \left(
        \begin{array}{cc}
        0 & 0 \\
        0 & h^{i j}
        \end{array}
        \right)
    \, .
\end{split}
\end{align}
Inversely, we can express the metric in terms of $N$, $N^{i}$, and $h_{ij}$,
\begin{align}
\begin{split}
    g_{\mu \nu} 
    &=\left(
        \begin{array}{cc}
        -N^{2}+N_{i} N^{i} & N_{i} \\
        N_{i} & h_{i j}
	\end{array}
        \right)
    \, , \\
    g^{\mu \nu} 
    &= \left(
        \begin{array}{cc}
	-N^{-2} & N^{-2} N^{i} \\
	N^{-2} N^{i} & h^{i j} - \frac{N^iN^j}{N^2}
	\end{array}
        \right)
    \, .
\end{split}
\label{Eq: ADM metric}
\end{align}
	
Moreover, we can decompose the space-time vector $A_{\mu}^{a}$ into tangent $\bar{A}_{\mu}^{a}$ and normal part $A_{*}^{a}$ separately:
\begin{align}
A_{\mu}^{a}=\bar{A}_{\mu}^{a}-A_{*}^{a}n_{\mu} \quad \text{with} \quad A_{*}^{a}=A_{\alpha}^{a}n^{\alpha} 
\, .
\end{align}
For the tangent part, we have $A_{0}^{a}=\bar{A}_{0}^{a}+A_{*}^{a}N$ and $A_{i}^{a}=\bar{A}_{i}^{a}$. Provided by these two relations, the normal component of $A_{\mu}^{a}$ can be expressed by
\begin{align}
\begin{split}
    A_{*}^{a}= & A_{0}^{a}\cdot n^{0}+A_{i}^{a}n^{i}
    \\
    & =
    \frac{1}{N}(\bar{A}_{0}^{a}+A_{*}^{a}N-N^{i}\bar{A}_{i}^{a})
    \, ,
\end{split}		
\end{align}
from which we get the relation $\bar{A}_{0}^{a}=  N^{i}\bar{A}_{i}^{a}$. 
After straightforward calculations presented in Appendix~\ref{appendix C}, one can obtain the different components of the covariant derivative of four vectors $A_{\mu}^{a}$:
\begin{align}
\begin{split}
    \nabla_{0}A_{0}^{a}
    & = 
    N\dot{A}_{*}^{a} - \mathcal{K}_{ij} \left( A_{*}^{a}N^{i}N^{j}+N\bar{A}^{ia}N^{j}+N\bar{A}^{ja}N^{i} \right)
    \\
    & - N^i A^a_\ast \mathcal{D}_i N - N^i \bar{A}^a_j \mathcal{D}_i N^j - N\bar{A}^{ia}\mathcal{D}_{i}N + N^{i} \dot{\bar{A}}_{i}^{a}
    \, , \\
    \nabla_{i}A_{0}^{a}
    & = 
    -\mathcal{K}_{ij} \left( A_{*}^{a}N^{j} + N\bar{A}^{ja} \right) + N\mathcal{D}_{i}A_{*}^{a} + N^{j}\mathcal{D}_{i}\bar{A}_{j}^{a}
    \, , \\
    \nabla_{0}A_{i}^{a}
    & =
    \dot{\bar{A}}_{i}^{a} - \mathcal{K}_{ij} \left(A_{*}^{a}N^{j}+N\bar{A}^{ja} \right)
    - A^a_\ast \mathcal{D}_i N - \bar{A}^a_j \mathcal{D}_i N^j
    \, , \\
    \nabla_{i}A_{j}^{a}
    & = 
    \mathcal{D}_{i}\bar{A}_{j}^{a}-A_{*}^{a}\mathcal{K}_{ij}
    \, .
\end{split}
\end{align}

Subsequently, by using the same approach, the $3+1$ decomposition of $B_{a}^{\alpha}$ is given by
\begin{align}
B_{a}^{0}=  -B_{a}^{*}\frac{1}{N}~,~
B_{a}^{i}=  \bar{B}_{a}^{i}+B_{a}^{*}\frac{N^{i}}{N}~,~\label{Eq.3+1 of B}
\end{align}
the calculation details of Eq.~\eqref{Eq.3+1 of B} and the $3+1$ decomposition of the interaction term $C_{a}^{\alpha\mu\nu}$ have also shown explicitly in Appendix~\ref{appendix C}. It is worth mentioning that since the $A_{a}^{\mu}$ and $B_{a}^{\mu}$ are not independent space-time vectors, they are related by the $B_{a}^{\mu}A_{\nu}^{a}=\delta^{\mu}_{\;\nu}$, which lead to the following important relations:
\begin{align}
A_{*}^{a}B_{a}^{*}=A_{\alpha}^{a}n^{\alpha}B_{a}^{\beta}n_{\beta}=\delta_{\;\alpha}^{\beta}n^{\alpha}n_{\beta}=-1~,~\label{Eq.relation A.B.1}
\end{align}
and 
\begin{align}
B_{a}^{\mu}A_{\mu}^{a}= B_{a}^{0}A_{0}^{a}+B_{a}^{i}A_{i}^{a}= \bar{B}_{a}^{i}\bar{A}_{i}^{a}-B_{a}^{*}A_{*}^{a}=4~,~\label{Eq.relation A.B.2}
\end{align}
from which, one concludes that the tangent parts of those vectors also enjoy the relation $\bar{B}_{a}^{i}\bar{A}_{j}^{a}=\delta^{i}_{j}$. In particular, we have
\begin{align}
\bar{A}_{j}^{a}B_{a}^{*}= A_{j}^{a}(B_{a}^{\alpha}n_{\alpha})=-\delta_{j}^{0}N=0~.~\label{Eq.relation A.B.3}
\end{align}
It shows the tangent part of $A_{a}^{\mu}$ and the normal part of $B_{a}^{\mu}$ are orthogonal to each other and vice versa, i.e., $A_{*}^{a}\bar{B}_{a}^{i}=0$. Putting together all our results and utilizing relation~\eqref{Eq.relation A.B.1}-\eqref{Eq.relation A.B.3}, we finally obtain the full $3+1$ decomposition of the action~\eqref{Eq.modified action},
\begin{align}
\begin{split}
    \mathcal{L}^{eq}_{f(Q)} 
    &=  \, \mathcal{L}_{f(R)} - \mathcal{L}_{\text{int}}+\omega_{a}^{\nu}(\partial_{\nu}\xi^{a}-A_{\nu}^{a}) 
    \\
    &= \, N \sqrt{h} \varphi \left( ^{(3)}\mathcal{R} + \mathcal{K}^{ij} \mathcal{K}_{ij} -\mathcal{K}^{2} - \frac{V(\varphi)}{\varphi} \right) 
    \\
    & \quad
    + \sqrt{h} \dot{\varphi}  \left( - \frac{1}{N} \bar{B}^i_a \dot{\bar{A}}^a_i + B^\ast_a \mathcal{D}_i \bar{A}^{ia} + \frac{N^i}{N} \bar{B}^j_a \mathcal{D}_i \bar{A}^a_j + \frac{1}{N} \mathcal{D}_i N^i \right) + \sqrt{h} \dot{A}^a_\ast \bar{B}^i_a \mathcal{D}_i \varphi 
    \\
    & \quad
    + \sqrt{h} \mathcal{D}_i \varphi \mathcal{D}^i N  + \sqrt{h} \frac{N^i}{N} \left( \bar{B}^j_a \dot{\bar{A}}^a_j - \mathcal{D}_j N^j - N^j \bar{B}^k_a \mathcal{D}_j \bar{A}^a_k \right) \mathcal{D}_i \varphi 
    \\
    & \quad
    - \sqrt{h} N \left(B^\ast_a \mathcal{D}^i A^a_\ast + \bar{B}^i_a \mathcal{D}^j \bar{A}^a_j - \bar{B}_{ja} \mathcal{D}^i \bar{A}^{ja}  \right) \mathcal{D}_i \varphi 
    \\
    & \quad
    - \sqrt{h} N^i \left(B^\ast_a \mathcal{D}_j \bar{A}^{ja} \mathcal{D}_i \varphi + \bar{B}^j_a \mathcal{D}_i A^a_\ast \mathcal{D}_j \varphi \right) 
    \\
    & \quad
    + \omega_{a}^{0}\left(\dot{\xi}^{a}-NA_{*}^{a}-N^{i}\bar{A}^{a}_{i}\right) + \omega_{a}^{i}(D_{i}\xi^{a}-\bar{A}_{i}^{a})
    \, ,
    \label{action.E.Q}
\end{split}
\end{align}
where $\mathcal{K}_{ij}$ is extrinsic curvature of the hypersurface,
\begin{align}
    \mathcal{K}_{i j}=\frac{1}{2 N}\left(\dot{h}_{i j}-D_i N_j-D_j N_i\right)
    \, .\label{definition.of.Kij}
\end{align}

In the Lagrangian formulation, the field equations obtained from~\eqref{action.E.Q} involve only first-order time derivatives of $\varphi$ and ${A}_{*}^{a}$. 
However, wave functions require second-order derivatives. In other words, the action~\eqref{action.E.Q} strongly indicates these scalar fields cannot propagate as physical DOF. 
Furthermore, we can glimpse the prospect of the Hamiltonian formulation. The canonical momentum variable with respect to the scalar field $\varphi$ is given as
\begin{align}
    p = \sqrt{h} \left[ - \frac{1}{N} \bar{B}^i_a \dot{\bar{A}}^a_i + B^\ast_a \mathcal{D}_i \bar{A}^{ia} + \frac{N^i}{N} \bar{B}^j_a \mathcal{D}_i \bar{A}^a_j  + \frac{1}{N} \mathcal{D}_i N^i  \right]
    \, .
    \label{Eq: conjugate momenta}
\end{align}
Moreover, we consider the constraint from the Lagrange multiplier $\omega^{\nu}_{\ a}$.
We obtain $\nabla_{\mu}A^{a}_{\ \nu} = \nabla_{\nu}A^{a}_{\ \mu}$ after we use the constraint $A^{a}_{\ \mu} = \partial_{\mu} \xi^{a}$~\cite{Langlois:2015skt}.
Since the dummy indices $\mu, \nu$ are symmetric, 
we find that $\nabla_{i}A_{0}^{a}$ equals to $\nabla_{0}A_{i}^{a}$, which induces a relation $\dot{\bar{A}}_{i}^{a}=\mathcal{D}_{i}A_{0}^{a}$.
If we use $\dot{\bar{A}}^a_i = \mathcal{D}_i A^a_0$ and $\mathcal{D}_i \bar{A}^a_j = \mathcal{D}_j \bar{A}^a_i$, Eq.~\eqref{Eq: conjugate momenta} is reduced to the following form:
\begin{align}
    p  \approx\sqrt{h}(B_{a}^{\ast}D_{i}\bar{A}^{ia}-\bar{B}_{a}^{i}D_{i}A_{*}^{a})~\label{Eq.p constraint}
    \, ,
\end{align}
which generates constraints between phase space variables on the hypersurfaces $\Sigma_{t}$.
In other words, we can eliminate the phase space variables $\varphi$ and its corresponding conjugate momentum $p$ by using Eq.~\eqref{Eq.p constraint} and other constraints. 
It indicates that the ghost scalar mode discussed in Sec.~\ref{sec3} is nonphysical; thus, we can conclude that the covariant $f(Q)$ gravity theory is free from the ghosts.

%%%%%%%%%%%%%%%%%%%%%%%%%
\section{Conclusions and discussions} \label{sec5}
%%%%%%%%%%%%%%%%%%%%%%%%%

In this work, we have revisited the theoretical structure of the $f(Q)$ gravity and discussed the ghost scalar mode in the physical DOF.
Four scalar fields $\xi^{a}$ in the affine connection play a role of the St\"{u}eckelberg to restore the diffeomorphism, similar to the dRGT massive gravity, and the coincident gauge in the nonmetricity gravity corresponds to the unitary gauge in the dRGT massive gravity~\cite{deRham:2014zqa,Hassan:2011tf}.
In the covariant $f(Q)$ gravity, we have considered the scalar-nonmetricity formulation and conformal re-scaling of the metric,
where this method was well established in $f(R)$ gravity to study the scalar field in the theory.
We have shown that the scalar field $\varphi$, which stems from the $f(Q)$ functional DOF, has the negative kinetic energy and is the ghost mode. This result is consistent with earlier work~\cite{BeltranJimenez:2021auj}, and the opposite sign in front of the kinetic term of the scalar field originates from the nonmetricity tensor when we rewrite the nonmetricity scalar by the Ricci scalar.

To further investigate the ghost mode, we have developed a new formulation of the $f(Q)$ gravity, 
inspired by the technique to address the higher derivative of the scalar field of the HOST theory.
Introducing the vector field $A^{a}_{\mu}$ to rewrite the derivative of the four St\"{u}eckelberg field $\partial_\mu \xi^{a}$,
we have shown that the covariant $f(Q)$ gravity written by the scalar field $\varphi$, vector fields $A^{a}_{\mu}$, and the tensor field $g_{\mu \nu}$ can be equivalent to the HOST theory. 
The structure of the corresponding Lagrangian is relatively simple and explicit, equals to $f(R)$ term plus an interaction term, which might be reminiscent of a special class of Horndeski theories in four dimensions, i.e., $L^{\text{H}}_4$~\cite{Horndeski:1974wa,Kobayashi:2019hrl}. 
However, it is worth mentioning that the nonquadratic interacting term in Lagrangian~\eqref{Eq.modified action} might indicate that the vectors $A^a_{\mu}$ are nonpropagate modes.
The coupling between the scalar and vector fields in the new formulation rephrases the opposite sign of the scalar field from the nonmetricity tensor.
Moreover, we have also applied $3+1$ decomposition to the new formulation of the $f(Q)$ gravity. 
By using ADM variables, we are able to write the full Lagrangian~\eqref{Eq.modified action} in a relatively compact form, which significantly simplifies the computation of the Hamiltonian for further study. 
At last, we found that in the context of canonical structure, the ghost scalar mode can be removed by using constraints; thus, it ceases to propagate, and the theory is free from the ghost.

We make several remarks on the ghost scalar field in $f(Q)$ gravity.
The ghost is a dynamical DOF with a negative norm and often cancels the physical DOF in quantum theory. 
Even in classical theory, if we consider the trace or determinant of the coefficients of the kinetic term, ghosts may lower the rank of the kinetic matrix. 
Therefore, if we incorrectly assume all the DOFs have the positive norm, the number of the DOFs looks different from what it really is.
It is mandatory to handle the Hamiltonian analysis and to evaluate the physical DOF, assuming that ghost mode can exist in the theory.
The new formulation we developed will be helpful to understand the theoretical structure of the covariant $f(Q)$ gravity.
In the future, it would be necessary to confirm the conjecture that whether it suffers from Ostrogradsky instability brought by higher-order derivatives~\cite{Kobayashi:2019hrl,Gleyzes:2014dya}, 
and we need to check the DOF of covariant $f(Q)$ gravity by performing complete Hamiltonian analysis.
Finally, we comment on the ghost mode at the quantum level.
Although we have confirmed that the ghost scalar is not propagating at the classical level, such a ghost might propagate at the quantum level. 
Nonpropagation of the ghost scalar at the classical level relies on the symmetry 
$\nabla_{\mu}A^{a}_{\ \nu} = \nabla_{\nu}A^{a}_{\ \mu}$ after we use the constraint $A^{a}_{\ \mu} = \partial_{\mu} \xi^a$.
It is intriguing to investigate the relation between the propagation of the ghost scalar and the symmetry induced by the constraint in terms of the quantum field theory.

\begin{acknowledgments}
T.K. is supported by the National Key R\&D Program of China (2021YFA0718500) and by Grant-in-Aid of Hubei Province Natural Science Foundation (2022CFB817). T.Q. and K.H. are supported by the National Key Research and Development Program of China under Grant No. 2021YFC2203100, and the National Science
Foundation of China under Grant No. 11875141. S.N. was partially supported by the Maria de Maeztu Visiting Professorship at the Institute of Space Sciences, Barcelona.
\end{acknowledgments}

\appendix

%%%%%%%%%%%%%%%%%%%%%%%%%
\section{THE GENERAL COORDINATE TRANSFORMATION: AN EXAMPLE} \label{appendix A} 
%%%%%%%%%%%%%%%%%%%%%%%%%
One can use Eq.~\eqref{Eq: general covariance} to freely transform from tangent space-time coordinates to arbitrary space-time coordinates. 
In order to better illustrate this point, we consider the nonmetricity tensor.
\begin{align}
\begin{split}
    Q_{\alpha\mu\nu} 
    & =
    \hat{\nabla}_{\alpha}g_{\mu\nu}
    = 
    \partial_{\alpha}g_{\mu\nu} - g_{\nu\sigma}\varGamma_{\;\mu\alpha}^{\sigma}
    -g_{\sigma\mu}\varGamma_{\;\nu\alpha}^{\sigma}
    \\
    & =
    \partial_{\alpha}g_{\mu\nu}
    -g_{\nu\sigma}\left(\frac{\partial x^{\sigma}}{\partial\xi^{c}}\frac{\partial^{2}\xi^{c}}{\partial x^{\mu}\partial x^{\alpha}}\right)
    -g_{\sigma\mu}\left(\frac{\partial x^{\sigma}}{\partial\xi^{c}}\frac{\partial^{2}\xi^{c}}{\partial x^{\nu}\partial x^{\alpha}}\right)
    \\
    & =
    \partial_{\alpha}g_{\mu\nu}
    -\left(\frac{\partial\xi^{a}}{\partial x^{\nu}}\frac{\partial\xi^{b}}{\partial x^{\sigma}}f_{ab}\right)\frac{\partial x^{\sigma}}{\partial\xi^{c}}
    \cdot\partial_{\alpha}\frac{\partial\xi^{c}}{\partial x^{\mu}}
    - \left(\frac{\partial\xi^{a}}{\partial x^{\sigma}}\frac{\partial\xi^{b}}{\partial x^{\mu}}f_{ab}\right)
    \left(\frac{\partial x^{\sigma}}{\partial\xi^{c}}\cdot\partial_{\alpha}\frac{\partial\xi^{c}}{\partial x^{\nu}}\right)
    \\
    & =
    \partial_{\alpha}g_{\mu\nu}
    - \frac{\partial\xi^{c}}{\partial x^{\nu}}f_{ac}\cdot\partial_{\alpha}\frac{\partial\xi^{a}}{\partial x^{\mu}}
    -\frac{\partial\xi^{a}}{\partial x^{\mu}}f_{ac}\cdot\partial_{\alpha}\frac{\partial\xi^{c}}{\partial x^{\nu}}
    \\
    & =
    \partial_{\alpha}g_{\mu\nu}
    - \left(
    \frac{\partial\xi^{c}}{\partial x^{\nu}}\cdot\partial_{\alpha}\frac{\partial\xi^{a}}{\partial x^{\mu}}
    + \frac{\partial\xi^{a}}{\partial x^{\mu}}\cdot\partial_{\alpha}\frac{\partial\xi^{c}}{\partial x^{\nu}}
    \right)f_{ac}
    \\
    & =
    \partial_{\alpha}g_{\mu\nu}
    - \partial_{\alpha}\left(\frac{\partial\xi^{c}}{\partial x^{\nu}}\frac{\partial\xi^{a}}{\partial x^{\mu}}f_{ac}\right)
    + \left(\partial_{\alpha}f_{ac}\right)\cdot\frac{\partial\xi^{c}}{\partial x^{\nu}}\frac{\partial\xi^{a}}{\partial x^{\mu}}
    \\
    & =
    \frac{\partial\xi^{c}}{\partial x^{\nu}}\frac{\partial\xi^{a}}{\partial x^{\mu}}\frac{\partial\xi^{d}}{\partial x^{\alpha}}\mathring{Q}_{dac} 
    \, .
\end{split}
\label{Eq.transformation fab}
\end{align}
We have used the condition $\mathring{Q}_{abc}=\partial_{a}f_{bc}$ in the last step. One may notice Eq.~\eqref{Eq.transformation fab} is nothing but the transformation of nonmetricity tensor between arbitrary gauge and coincident gauge.

%%%%%%%%%%%%%%%%%%%%%%%%%
\section{SCALAR-VECTOR-TENSOR REPRESENTATION OF $f(Q)$ GRAVITY} \label{appendix B} 
%%%%%%%%%%%%%%%%%%%%%%%%%

We show the calculation detail in Eq.~\eqref{Eq.modified action} up to $S_{f(R)}$ and Lagrange multiplier $\omega_{\;a}^{\nu}(\partial_{\nu}\xi^{a}-A_{\;\nu}^{a})$:
\begin{align}
    \int d^{4}x\, \left[\sqrt{-g}\partial_{\alpha}\varphi\cdot(Q^{\alpha}-\tilde{Q}^{\alpha}) \right]
    \, .
\end{align}
Computing the integrand, we find the traces of the nonmetricity tensor are written by metric $g_{\mu\nu}$ and vector field $A^{a}_{\ \mu}$ as follows:
\begin{align}
    Q^{\alpha}-\tilde{Q}^{\alpha} 
    &= 
    g^{\beta\alpha}g^{\mu\nu}(\partial_{\beta}g_{\mu\nu}-\partial_{\mu}g_{\beta\nu})
    + \left[ g^{\mu\nu}(A^{-1})_{\;a}^{\alpha}-g^{\mu\alpha}(A^{-1})_{\;a}^{\nu} \right] \partial_{\mu}A_{\;\nu}^{a}
    \, .
    \label{Eq: boundary a1}
\end{align}
Using the following two relations
\begin{align}
\begin{split}
    \partial_{\beta}g_{\mu\nu}
    &=
    g_{\lambda \nu} \left\{ {}^{\, \lambda}_{\mu\beta} \right\} + g_{\lambda\mu} \left\{ {}^{\, \lambda}_{\nu\beta} \right\}
    \, , \\
    \nabla_{\mu}A_{\nu}^{a}
    &=
    \partial_{\mu}A_{\nu}^{a} - \left\{ {}^{\, \alpha}_{\mu\nu} \right\} A_{\alpha}^{a}
    \, ,
\end{split}
\end{align}
we can compute the first term in Eq.~\eqref{Eq: boundary a1} as
\begin{align}
\begin{split}
    g^{\beta\alpha}g^{\mu\nu} \left( \partial_{\beta}g_{\mu\nu}-\partial_{\mu}g_{\beta\nu} \right)
    & =
    \left( g^{\beta\alpha}g^{\mu\nu} - g^{\mu\alpha}g^{\beta\nu} \right) \partial_{\beta}g_{\mu\nu}
    \\
    & =
    \left( g^{\beta\alpha}g^{\mu\nu} - g^{\mu\alpha}g^{\beta\nu} \right) 
    \left(
    g_{\lambda\nu} \left\{{}^{\, \lambda}_{\mu\beta} \right\} + g_{\lambda\mu} \left\{{}^{\, \lambda}_{\nu\beta} \right\} 
    \right)
    \\
    &=
    g^{\beta\alpha} \left\{{}^{\, \lambda}_{\lambda\beta} \right\} - g^{\beta\nu} \left\{{}^{\, \alpha}_{\nu\beta} \right\}~,
\end{split}
\end{align}
and the second term in Eq.~\eqref{Eq: boundary a1} as
\begin{align}
\begin{split}
    \left[ g^{\mu\nu}(A^{-1})_{\, a}^{\alpha} - g^{\mu\alpha}(A^{-1})_{\, a}^{\nu} \right] \cdot\partial_{\mu}A_{\;\nu}^{a}
    & =
    \left[ g^{\mu\nu}(A^{-1})_{\, a}^{\alpha} - g^{\mu\alpha}(A^{-1})_{\, a}^{\nu} \right]
    \cdot \left( \nabla_{\mu}A_{\;\nu}^{a} + \left\{{}^{\, \lambda}_{\mu\nu} \right\}A_{\, \lambda}^{a} \right)
    \\
    & =
    \left[
    g^{\mu\nu}(A^{-1})_{\;a}^{\alpha}\cdot \left\{{}^{\, \lambda}_{\mu\nu} \right\}A_{\, \lambda}^{a} 
    - g^{\mu\alpha}(A^{-1})_{\;a}^{\nu} \cdot \left\{ {}^{\, \lambda}_{\mu\nu} \right\}A_{\, \lambda}^{a}
    \right]
    \\
    & \qquad 
    + \left[g^{\mu\nu}(A^{-1})_{\;a}^{\alpha}-g^{\mu\alpha}(A^{-1})_{\;a}^{\nu} \right] \cdot \nabla_{\mu}A_{\;\nu}^{a}
    \\
    & =
    g^{\mu\nu} \cdot \left\{{}^{\, \alpha}_{\mu\nu} \right\} - g^{\mu\alpha} \cdot \left\{{}^{\, \nu}_{\mu\nu} \right\}
    \\
    & \qquad 
    + \left[g^{\mu\nu}(A^{-1})_{\;a}^{\alpha}-g^{\mu\alpha}(A^{-1})_{\;a}^{\nu} \right] \cdot \nabla_{\mu}A_{\;\nu}^{a}~.
\end{split}
\end{align}
% and with the help of relation~\eqref{relation.Q.R}, we calculate
Finally, we obtain
\begin{align}
    Q^{\alpha}-\tilde{Q}^{\alpha} 
    &= 
    \left[g^{\mu\nu}(A^{-1})_{\;a}^{\alpha}-g^{\mu\alpha}(A^{-1})_{\;a}^{\nu} \right] \cdot \nabla_{\mu}A_{\;\nu}^{a}~,
    \label{Eq: boundary a2}
\end{align}
and thus
\begin{align}
\begin{split}
    \int d^{4}x\, \left[\sqrt{-g}\partial_{\alpha}\varphi\cdot(Q^{\alpha}-\tilde{Q}^{\alpha}) \right]
    = 
    \int d^{4}x\, \left[\sqrt{-g} \partial_{\alpha}\varphi \left[g^{\mu\nu}(A^{-1})_{\;a}^{\alpha}-g^{\mu\alpha}(A^{-1})_{\;a}^{\nu} \right] \cdot \nabla_{\mu}A_{\;\nu}^{a} \right]
    \, .
\end{split}
\end{align}

%%%%%%%%%%%%%%%%%%%%%%%%%
\section{ADM DECOMPOSITION OF THE ACTION} \label{appendix C} 
%%%%%%%%%%%%%%%%%%%%%%%%%
This Appendix aims to apply ADM decomposition to the action~\eqref{Eq.modified action}. In order to facilitate the calculation, we will divide our goal into two parts $\nabla_{\mu}A_{\nu}^{a}$, $C_{a}^{\alpha\mu\nu}$, and compute them separately.

%%%%%%%%%%%%%%%%%%%%%%%%%
\subsubsection{$3+1$ decomposition of $\nabla_{\mu}A_{\nu}^{a}$}
%%%%%%%%%%%%%%%%%%%%%%%%%

Utilizing the expressions of the Christoffel symbols~\eqref{Levi-Civita connection} in terms of ADM quantities~\eqref{Eq: ADM metric}, we have
\begin{align}
\begin{split}
    \left\{{}^{\, 0}_{00} \right\} 
    & = 
    \frac{1}{N} \left( N^{i}N^{j}\mathcal{K}_{ij}+\dot{N}+N^{i}\mathcal{D}_{i}N \right)
    \, , \\
    \left\{{}^{\, k}_{00} \right\} 
    & = 
    NN^{i}\left(2h^{jk}-\frac{N^{j}N^{k}}{N^{2}}\right)\mathcal{K}_{ij}+\dot{N^{k}}
    \\
    & \qquad
    - \frac{N^{k}}{N}\dot{N} + N^{i}\mathcal{D}_{i}N^{k} + N \left(h^{ki} - \frac{N^{k}N^{i}}{N^{2}}\right) \mathcal{D}_{i}N
    \, , \\
    \left\{{}^{\, 0}_{0i} \right\} 
    & =
    \frac{1}{N} \left( N^{j} \mathcal{K}_{ij} + \mathcal{D}_{i}N \right)
    \, , \\
    \left\{{}^{\, j}_{0i} \right\}  
    & = 
    N \left(h^{jk} - \frac{N^{j}N^{k}}{N^{2}}\right)\mathcal{K}_{ik} + \mathcal{D}_{i}N^{j} - \frac{N^{j}}{N}\mathcal{D}_{i}N
    \, , \\
    \left\{{}^{\, 0}_{ij} \right\}  
    & =
    \frac{1}{N}\mathcal{K}_{ij}
    \, , \\
    \left\{{}^{\, k}_{ij} \right\}  
    & =
    -\frac{N^{k}}{N}\mathcal{K}_{ij} + ^{3}\left\{{}^{\, k}_{ij} \right\} 
    \, .     
\end{split}
\end{align}
where $\mathcal{K}_{ij}$ is extrinsic curvature of the hypersurface defined by Eq.~\eqref{definition.of.Kij}. From these expressions, we compute each component of the covariant derivative of vector $\nabla_{\mu}A_{\nu}^{a}$:
\begin{align}
\begin{split}
    \nabla_{0}A_{0}^{a}
    & = 
    N\dot{A}_{*}^{a} - \mathcal{K}_{ij} \left( A_{*}^{a}N^{i}N^{j}+N\bar{A}^{ia}N^{j}+N\bar{A}^{ja}N^{i} \right)
    \\
    & - N^i A^a_\ast \mathcal{D}_i N - N^i \bar{A}^a_j \mathcal{D}_i N^j - N\bar{A}^{ia}\mathcal{D}_{i}N + N^{i} \dot{\bar{A}}_{i}^{a}
    \, , \\
    \nabla_{i}A_{0}^{a}
    & = 
    -\mathcal{K}_{ij} \left( A_{*}^{a}N^{j} + N\bar{A}^{ja} \right) + N\mathcal{D}_{i}A_{*}^{a} + N^{j}\mathcal{D}_{i}\bar{A}_{j}^{a}
    \, , \\
    \nabla_{0}A_{i}^{a}
    & =
    \dot{\bar{A}}_{i}^{a} - \mathcal{K}_{ij} \left(A_{*}^{a}N^{j}+N\bar{A}^{ja} \right)
    - A^a_\ast \mathcal{D}_i N - \bar{A}^a_j \mathcal{D}_i N^j
    \, , \\
    \nabla_{i}A_{j}^{a}
    & = 
    \mathcal{D}_{i}\bar{A}_{j}^{a}-A_{*}^{a}\mathcal{K}_{ij}
    \, .
\end{split}
\end{align}

%%%%%%%%%%%%%%%%%%%%%%%%%
\subsubsection{$3+1$ decomposition of $B_{a}^{\alpha}$ and $C_{a}^{\alpha\mu\nu}$}
%%%%%%%%%%%%%%%%%%%%%%%%%

First, we apply $3+1$ decomposition to vector field $B_{a}^{\alpha}$.
From the property $B_{a}^{\alpha}= \bar{B}_{a}^{\alpha}-B_{a}^{*}n^{\alpha}$, we find
\begin{align}
\begin{split}
    B_{a}^{0}= & \bar{B}_{a}^{0}-B_{a}^{*}\frac{1}{N}
    \, , \\
    B_{a}^{i}= & \bar{B}_{a}^{i}+B_{a}^{*}\frac{N^{i}}{N}
    \, .\\
\end{split}
\end{align}
The norm part of the vector $B_{a}^{\alpha}$ is given as
\begin{align}
\begin{split}
    B_{a}^{*} 
    & = 
    B_{a}^{\alpha}n_{\alpha}=  B_{a}^{0}n_{0}
    \\
    & = 
    - N \left(\bar{B}_{a}^{0}-B_{*}\frac{1}{N}\right)
    \\
    &= 
    - N\bar{B}_{a}^{0}+B_{a}^{*}
    \, , 
\end{split}
\end{align}
from which we can conclude $\bar{B}_{a}^{0} = 0$. 
Thus, we obtain
\begin{align}
    B_{a}^{0}=  -B_{a}^{*}\frac{1}{N}
    \, , \ 
    B_{a}^{i}=  \bar{B}_{a}^{i}+B_{a}^{*}\frac{N^{i}}{N}
    \, .
    \label{Eq.B}
\end{align}

Inserting Eq.~\eqref{Eq.B} into Eq.~\eqref{Eq.tensor C}, we can compute each component of $C_{a}^{\alpha\mu\nu}$:
\begin{align}
\begin{split}
    C^{000}_a 
    &= g^{00} B^0_a - g^{00} B^0_a = 0 \, ,
    \\
    C^{00i}_a 
    &= g^{0i} B^0_a - g^{00} B^i_a = \frac{N^i}{N^2} B^0_a + \frac{1}{N^2} B^i_a = \frac{1}{N^2} \bar{B}^i_a \, ,
    \\
    C^{0i0}_a 
    &= g^{i0} B^0_a - g^{i0} B^0_a = 0 \, ,
    \\
    C^{i00}_a 
    &= g^{00} B^i_a - g^{0i} B^0_a = - \frac{N^i}{N^2} B^0_a - \frac{1}{N^2} B^i_a = - \frac{1}{N^2} \bar{B}^i_a \, ,
    \\
    C^{ij0}_a 
    &= g^{j0} B^i_a - g^{ji} B^0_a = - \left(h^{ij} - \frac{N^i N^j}{N^2} \right) B^0_a + \frac{N^j}{N^2} B^i_a = \frac{h^{ij}}{N} B^\ast_a + \frac{N^j}{N^2} \bar{B}^i_a \, ,
    \\
    C^{i0j}_a 
    &= g^{0j} B^i_a - g^{0i} B^j_a = \frac{N^j}{N^2} B^i_a - \frac{N^i}{N^2} B^j_a = \frac{1}{N^2} \left(N^j \bar{B}^i_a - N^i \bar{B}^j_a \right) \, ,
    \\
    C^{0ij}_a 
    &= g^{ij} B^0_a - g^{i0} B^j_a = \left(h^{ij} - \frac{N^i N^j}{N^2} \right) B^0_a - \frac{N^i}{N^2} B^j_a = - \frac{h^{ij}}{N} B^\ast_a - \frac{N^i}{N^2} \bar{B}^j_a  \, ,
    \\
    C^{ijk}_a 
    &= g^{jk} B^i_a - g^{ji} B^k_a = \left(h^{jk} - \frac{N^j N^k}{N^2} \right) B^i_a - \left(h^{ij} - \frac{N^i N^j}{N^2} \right) B^k_a 
    \\
    &= \frac{1}{N} \left(h^{jk} N^i - h^{ij} N^k \right) B^\ast_a + \left(h^{jk} - \frac{N^j N^k}{N^2} \right) \bar{B}^i_a - \left(h^{ij} - \frac{N^i N^j}{N^2} \right) \bar{B}^k_a \, .
\end{split}
\end{align}

\bibliographystyle{apsrev4-1}
\bibliography{References.bib}

\end{document}